\documentclass{elsart}
\usepackage{natbib}
\def\aprle{\buildrel < \over {_{\sim}}}

\begin{document}
\hfill FERMILAB-Conf-01/120-T
\runauthor{Albright}
\begin{frontmatter}
\title{GUT implications from neutrino mass\thanksref{X}}
\vspace*{-0.2in}
\author{Carl H. Albright\thanksref{Y}\thanksref{Z}}\\[-0.1in]
\address{Fermi National Accelerator Laboratory, Batavia, IL 60510\\[-0.3in]}
\thanks[X]{Invited talk at the NuFACT'01 Workshop in Tsukuba,
	Japan, 24-30 May 2001.}
\thanks[Y]{electronic address: albright@fnal.gov}
\thanks[Z]{Supported in part by the U.S. Department of Energy}
\begin{abstract}
An overview is given of the experimental neutrino mixing results and types of 
neutrino models proposed, with special attention to the general features of 
various GUT models involving intra-family symmetries and horizontal flavor 
symmetries.  Many of the features are then illustrated by a specific 
$SO(10)$ SUSY GUT model formulated by S.M. Barr and the author which can 
explain all four types of solar neutrino mixing solutions by various choices 
of the right-handed Majorana mass matrix.  The quantitative nature of the 
model's large mixing angle solution is used to compare the reaches of a 
neutrino super beam and a neutrino factory for determining the small 
$U_{e3}$ mixing matrix element.  
\end{abstract}
\end{frontmatter}
\vspace*{-0.4in}
\section{Introduction}
\vspace*{-0.2in}
\subsection{Summary of neutrino mass and mixing data}
\vspace*{-0.1in}
With $\Delta m^2_{ij} \equiv m^2_i - m^2_j$ and $U_{\ell i}$, one of the 
Maki-Nakagawa-Sakata (MNS) neutrino mixing matrix elements \cite{mns}, the 
following information is known.
\begin{itemize}
	\item 	{\bf Atmospheric Neutrinos}\\
		Interpretation of the data heavily favors $\nu_\mu 
		\leftrightarrow\nu_\tau$ over $\nu_\mu \leftrightarrow 
		\nu_{\rm sterile}$ oscillations \cite{atm} with oscillation 
		parameters
		\begin{equation}
		\begin{array}{rl}
		\Delta m^2_{32} &\simeq 3.2 \times 10^{-3}\ {\rm eV^2},\\
  		\sin^2 2\theta_{\rm atm} &= 4|U_{\mu 3}|^2
                	|U_{\tau 3}|^2 = 1.000,\ (> 0.89\ @\ 90\%\ {\rm c.l.})
			\\
		\end{array}
		\end{equation}
	\item	{\bf Solar Neutrinos}\\
		Here the data favors $\nu_e \leftrightarrow \nu_\mu,\ \nu_\tau$
		oscillations over $\nu_e \leftrightarrow \nu_{\rm sterile}$
		oscillations, but four solar mixing solutions were possible
		as reported by Gonzalez-Garcia at last year's Osaka HEP
		International Conference \cite{g-g}; i.e., the small mixing 
		angle (SMA), the large mixing angle (LMA), and the (LOW) MSW 
		\cite{msw} solutions, together with the quasi-vacuum 
		(QVO) solution.  In the (3 active, 0 sterile) neutrino 
		framework, the best-fit mixing solutions obtained with
		$\sin^2 2\theta_{sol} = 4|U_{e1}|^2|U_{e2}|^2$ were:
		\begin{equation}
		\begin{array}{rll}
		&SMA: &\Delta m^2_{21} \simeq 5.0 \times 10^{-6}\ {\rm eV^2},\\
		&	&\sin^2 2\theta_{sol} \simeq 0.0024,\ \tan^2 \theta
			 	\simeq 0.0006;\\
		&LMA:  &\Delta m^2_{21} \simeq 3.2 \times 10^{-5}\ {\rm eV^2},
			\\
		&	&\sin^2 2\theta_{sol} \simeq 0.75,\ \tan^2 \theta
			 	\simeq 0.33;\\
		&LOW:   &\Delta m^2_{21} \simeq 1.0 \times 10^{-7}\ 
				{\rm eV^2},\\
		&	&\sin^2 2\theta_{sol} \simeq 0.96,\ \tan^2 \theta
			 	\simeq 0.67;\\
		&QVO:   &\Delta m^2_{21} \simeq 8.6 \times 10^{-10}\ 
			{\rm eV^2},\\
		&	&\sin^2 2\theta_{sol} \simeq 0.96,\ \tan^2 \theta
			 	\simeq 1.5.\\
		\end{array}
		\end{equation}
		A new analysis by the Super-Kamiokande collaboration based 
		on 1258 days of data \cite{sol} now indicates that the LMA 
		region is strongly preferred over the other regions with the 
		best-fit point given by 
		\begin{equation}
		\begin{array}{rll}
		&LMA:  &\Delta m^2_{21} \simeq 7 \times 10^{-5}\ {\rm eV^2},\\
		&	&\sin^2 2\theta_{sol} \simeq 0.87,\ \tan^2 \theta
			 	\simeq 0.47.\\
		\end{array}
		\end{equation}
	\end{itemize}
With just three active neutrinos and maximal atmospheric mixing, to a good 
approximation one can simplify the general MNS mixing matrix to read
	\begin{equation}
		U_{MNS} \simeq \left(\matrix{c_{12} & s_{12} & 0\cr
		  -s_{12}/\sqrt{2} & c_{12}/\sqrt{2} & 1/\sqrt{2}\cr
		  s_{12}/\sqrt{2} & -c_{12}/\sqrt{2} & 1/\sqrt{2}\cr}
		  \right),\\
	\end{equation}
where the mass eigenstates are given in terms of the flavor states by 
	\begin{equation}
	\begin{array}{rl}
		\nu_3 &= \frac{1}{\sqrt{2}}(\nu_\mu + \nu_\tau),\\
		\nu_2 &= \nu_e s_{12} + \frac{1}{\sqrt{2}}(\nu_\mu - \nu_\tau)
			c_{12},\\
		\nu_1 &= \nu_e c_{12} - \frac{1}{\sqrt{2}}(\nu_\mu - \nu_\tau)
			s_{12}.\\
	\end{array}
	\end{equation}

In the LMA case with the latest data given above, one then finds
	\begin{equation}
		U^{(LMA)}_{MNS} \simeq \left(\matrix{0.825 & 0.565 & 0\cr
			-0.400 & 0.583 & 0.707\cr 
			0.583 & -0.400 & 0.707\cr}\right).
	\end{equation}
Note that with maximal atmospheric mixing, unitarity forces $\theta_{13} = 
0^o$; in fact, a primary physics goal of a neutrino factory is to determine 
$U_{e3}$ and its departure from zero.  While solar neutrino mixing is close 
to maximal, strictly maximal mixing is presently excluded at the 95\% c.l.  
If this persists, it may result in a severe test for models to predict the 
deviation from maximal mixing.
\vspace*{-0.2in}
\subsection{Types of neutrino models}
\vspace*{-0.3in}
The general neutrino mass matrix in a basis with $n_L$ left-handed fields
and $n_R = n^c_L$ right-handed or left-handed conjugate fields has the 
complex symmetric form
	\begin{equation}
	{\bf M}_\nu = \left(\matrix{M_L & N^T\cr N & M_R\cr}\right),
	\end{equation}
where $M_L$ is the left-handed Majorana mass matrix, $N$ and $N^T$ the 
Dirac mass matrix and its transpose, and $M_R$ is the right-handed Majorana
mass matrix.  Models which appear in the literature\footnote{More complete
surveys can be found in \cite{lit}.} can generally 
be placed into three classes as follows:
\begin{itemize}
\item	{\bf Models with only left-handed neutrinos present}\\
	Models of this type are variations of the Zee model \cite{zee}, where
	ultralight neutrinos arise from non-renormalizable contributions
	involving some undetermined high mass scale.  Lepton number is 
	violated by two units, or an $L = -2$ isovector Higgs field is 
	introduced.  The combination $L' \equiv L_e - L_\mu - L_\tau$
	is often taken to be conserved.
\item	{\bf Models with both left- and right-handed neutrinos present}\\
	With $M_L = 0$, the seesaw mechanism yields ultralight 
	neutrino masses provided the right-handed masses are in the range of 
	$10^5 - 10^{15}$ GeV.  Such masses are naturally obtained in GUT 
	models with $\Lambda_{GUT} = 2 \times 10^{16}$ GeV.
\item	{\bf Models with neutrinos in higher dimensions}\\
	Right-handed neutrinos which are singlets under all gauge symmetries
	can enter the bulk with many Kalusa-Klein states present.  With 
	large extra dimensions and the compactification scale much lower
	than the string scale, a modified seesaw mechanism can generate
	ultralight neutrino masses \cite{dien}.
\end{itemize}
\vspace*{-0.3in}
\section{Features of various GUT models}
\vspace*{-0.2in}
Restricting our attention to the second class of models, we note the 
intra-family symmetry specified by a GUT model provides a unified 
treatment of quarks and leptons as (some) quarks and leptons are placed
in the same multiplets.  For example, the representation content of
three familiar GUTs is listed below:\\[-0.2in]
$$\begin{array}{rl}
          SU(5):\ & (u_\alpha,\ d_\alpha,\ u^c_\alpha,\ \ell^c)_i\ \subset
                {\bf 10}_i\\
                & (d^c_\alpha,\ \ell,\ \nu_\ell)_i\ \subset 
                    {\bf \overline{5}}_i,\quad \alpha = r,b,g;\ 
                        i = 1,2,3\\
                & (\nu^c_\ell)_i\ \subset {\bf 1}_i\\[0.1in]
          SO(10): & (u_\alpha,\ d_\alpha,\ u^c_\alpha,\ d^c_\alpha,\ \ell,
                \ \ell^c,\ \nu_\ell,\ \nu^c_\ell)_i\ \subset {\bf 16}_i
		\\[0.1in]
          E_6:  & (u_\alpha,\ d_\alpha,\ u^c_\alpha,\ d^c_\alpha,\ \ell,
                \ \ell^c,\ \nu_\ell,\ \nu^c_\ell)_i,\\
		& (D_\alpha,\ D^c_\alpha,\ E,\ E^c,\ N,\ N^c)_i,\ 
                  n_i\ \subset {\bf 27}_i\\
\end{array}$$
\vspace*{-0.5in}

\noindent Sterile neutrinos can appear non-trivially in $E_6$ or as isolated 
singlets in $SU(5)$ or $SO(10)$.

On the other hand, a specified horizontal flavor symmetry enables one to 
connect comparable flavors in different families which allows a mass 
hierarchy to exist among the families.  The flavor symmetry may be discrete, 
such as $Z_2,\ S_3,\ Z_2 \times Z_2$ etc., and results in multiplicative 
quantum numbers.  A continuous flavor symmetry, such as $U(1),\ U(2),\ 
SU(3)$ etc., results in additive quantum numbers and may be global or local 
(and possibly anomalous).  

With a GUT family symmetry, some or all of the flavor bases are related for 
the up and down quark, charged lepton and neutrino Dirac mass matrices, 
hereafter denoted by $U,\ D,\ L,$ and $N$.  As such, the Yukawa interactions 
in GUT models are typically not diagonal in flavor space.  Contrast this with 
models with no grand unification, where the quark and lepton sectors can be 
treated independently and some matrices can be arbitrarily assumed diagonal.

In unbroken $SU(5)$, $L = D^T$, but $N$ and $D$ are unrelated, while the 
right-handed Majorana mass matrix, $M_R$ may or may not exist.  This tends
to provide a lot of freedom for the model builder as different flavor charges
can be assigned to each ${\bf 10,\ \overline{5}}$ and ${\bf 1}$ in the same
family.

In unbroken $SO(10)$, $U = D = L = N$ as all left-handed quarks and leptons 
belonging to the same family have the same flavor charge, while $M_R$ exists 
and is independent of the others due to its different Higgs VEV structure.
With $SU(5)$ and $SO(10)$ broken at the GUT scale, and the Higgs
fields as well as the fermion fields carrying horizontal flavor
quantum numbers, a rather complex set of mass matrix textures can emerge.

In $E_6$ eleven extra states are present in each fundamental ${\bf 27}$ which 
must be made heavy, aside possibly from 1 or 2 light sterile neutrinos
per family.
\vspace*{-0.3in}
\section{Symmetry breaking in $SO(10)$}
\vspace*{-0.2in}
Now restricting our attention to $SO(10)$, we note that to break $SO(10)$ to 
the SM, the rank must be reduced from 5 to 4 typically along one of the chains:
$$\begin{array}{rl}
        SO(10) & \rightarrow SU(5) \times U(1) \rightarrow SU(5) \rightarrow 
		{\rm SM},\\[0.1in]
        SO(10) & \rightarrow SU(4) \times SU(2)_L \times SU(2)_R\\
                &\rightarrow SU(3) \times SU(2)_L \times SU(2)_R \rightarrow 
                {\rm SM}\\
\end{array}$$
Among the possible $SO(10)$ Higgs VEVs in the first breaking chain at the 
GUT scale are:\\
\hspace*{0.5in} $\langle {\bf 45_H} \rangle$ \begin{minipage}[t]{4.5in}{which 
		can point in the $I_{3R},\ Y$ or $B - L$ direction, but these 
		do not reduce the rank;}\end{minipage}\\[0.1in]
\hspace*{0.5in} $\langle 1({\bf 16_H})\rangle$ which 
		breaks $SO(10) \rightarrow SU(5)$ by reducing the rank;
		\\[0.1in]
\hspace*{0.5in} $\langle {\bf 45_H}\rangle_{B - L} + \langle 1({\bf 16_H})
		\rangle$ which breaks $SO(10) \rightarrow 
		{\rm SM}$.

As for the electroweak Higgs doublets, they can appear in the 
$5$ and $\overline{5}$ representations of $SU(5)$ and will break 
${\rm SM} \rightarrow SU(3)_c \times U(1)_{em}$ at the electroweak scale,
provided they remain light while the Higgs color triplets get massive at 
$\Lambda_{GUT}$.  This is known as the doublet-triplet splitting problem.
In the standard procedure, the $5$ and $\overline{5}$ are placed in the 
same ${\bf 10_H}$ of $SO(10)$ enabling Yukawa coupling unification with 
$\tan \beta \simeq 55$.  However, another possibility is to place the
$5$ in the ${\bf 10_H}$ while the $\overline{5}$ belongs to a linear 
combination of the ${\bf 10_H}$ and a ${\bf 16_H}$.  This enables Yukawa 
unification with $1 \aprle \tan \beta \aprle 55$.  In any case, only two 
Higgs doublets can survive down to the electroweak scale for proper gauge 
coupling unification with $\sin^2 \theta_W \sim 0.2315$.

GUT models then differ by their choice of unification group, symmetry-breaking
schemes and assigned flavor symmetries.  Among $SO(10)$ GUT models, the 
following flavor symmetries appear in the literature:
	$$\begin{array}{ll}
                U(1)    & \quad {\rm Babu,\ Pati,\ Wilczek}\ (9)\cr
                U(1) \times Z_2 \times Z_2      & \quad 
                        {\rm Albright,\ Babu,\ Barr}\ (10)\cr
                SU(2) \times Z_2 \times Z_2 \times Z_2 & \quad 
                        {\rm Chen,\ Mahanthappa}\ (11)\cr
                U(2) \times U(1)^n & \quad {\rm Blazek,\ Raby,\ Toby}\ 
			(12)\cr
                SU(3) & \quad {\rm Berezhiani,\ Rossi}\ (13)\ \cr
        \end{array}$$
All rely on the seesaw mechanism \cite{gmrsy}, $M_\nu = - N^T M_R N$,
to obtain the light effective LH Majorana mass matrix, but $M_R$ may be 
generated with a $\langle 1({\bf \overline{126}_H})\rangle$ or a pair of 
$\langle 1({\bf \overline{16}_H})\rangle$'s.  With $U_{MNS} = 
U^\dagger_L U_\nu$, all models generate the maximal atmospheric $\nu_\mu - 
\nu_\tau$ mixing either by a special feature of $N$, a special feature of 
$L$, or the combined effect of $N$ and $M_R$ in the seesaw mechanism.
Most models easily accommodate the SMA solar solution, while 
some can accommodate the QVO or LOW solution as well.  However, most have 
great difficulty with, or find it impossible to explain, the LMA 
solution, since fine tuning is required.  This is especially true of 
models which require special features of $N$ and/or $M_R$ 
to get maximal atmospheric mixing.
\vspace*{-0.2in}
\section{$SO(10)$ SUSY GUT model with $U(1) \times Z_2 \times Z_2$ 
	flavor symmetry}
\vspace*{-0.2in}
We now illustrate a model, developed in collaboration with S.M. Barr
\cite{ab}, which is particularly
useful in that it is quantitatively predictive, can explain the LMA
solution, and can be used to assess the need for a neutrino 
factory.

It is based on a minimum set of Higgs fields which solves the doublet-triplet
splitting problem with just one ${\bf 45}_H$ whose VEV points in the 
$B-L$ direction and has no higher rank representations.  Two pairs of 
${\bf 16}_H,\ {\bf \overline{16}}_H$'s stabilize the solution \cite{br}.  
Several Higgs in the ${\bf 10}_H$ representations together with Higgs singlets
are also present.  The combination of VEVs, $\langle {\bf 45_H}\rangle_{B-L},
\ \langle 1({\bf 16_H})\rangle$ and $\ \langle 1({\bf \overline{16}_H})\rangle$
break $SO(10)$ to the SM.  The electroweak VEVs arise from $v_u = 
\langle 5({\bf 10_H})\rangle$ and $v_d = \langle \overline{5}({\bf 10_H})
\rangle\cos \gamma + \langle \overline{5}({\bf 16'_H})\rangle \sin \gamma$,
while the combination orthogonal to $v_d$ gets massive at the GUT scale.
The Higgs superpotential in this model exhibits the $U(1) \times Z_2 \times 
Z_2$ flavor symmetry.  

In addition, matter superfields appear in the following representations:\\
${\bf 16_1},\ {\bf 16_2},\ {\bf 16_3};\ {\bf 16},\ {\bf \overline{16}},
\ {\bf 16'},\ {\bf \overline{16'}},\ {\bf 10_1},\ {\bf 10_2}$, and ${\bf 1}$'s,
where all but the ${\bf 16_i},\ i = 1,2,3$ get superheavy and are integrated 
out.

The Dirac mass matrices are found to be\\[-0.2in] 
\begin{equation}
\begin{array}{ll}
U = \left(\matrix{ \eta & 0 & 0 \cr
  0 & 0 & \epsilon/3 \cr 0 & - \epsilon/3 & 1\cr} \right)M_U,\ 
  & D = \left(\matrix{ 0 & \delta & \delta' e^{i\phi}\cr
  \delta & 0 & \sigma + \epsilon/3  \cr
  \delta' e^{i \phi} & - \epsilon/3 & 1\cr} \right)M_D, \\[0.3in]
N = \left(\matrix{ \eta & 0 & 0 \cr 0 & 0 & - \epsilon \cr 
	0 & \epsilon & 1\cr} \right)M_U,\ 
  & L = \left(\matrix{ 0 & \delta & \delta' e^{i \phi} \cr
  \delta & 0 & -\epsilon \cr \delta' e^{i\phi} & 
  \sigma + \epsilon & 1\cr} \right)M_D,\\
\end{array}
\end{equation}
\vspace*{-0.2in}
where
\vspace*{-0.2in}
\begin{equation}
\begin{array}{rlrl}
        M_U&\simeq 113\ {\rm GeV},&\qquad M_D&\simeq 1\ {\rm GeV},\\
        \sigma&=1.78,&\qquad \epsilon&=0.145,\\
        \delta&=0.0086,&\qquad \delta'&= 0.0079,\\
        \phi&= 54^o,&\qquad \eta&= 8 \times 10^{-6}\\
\end{array}
\end{equation}
are input parameters defined at the GUT scale to fit the low scale 
observables after evolution downward from $\Lambda_{GUT}$.
  
The above textures were obtained by imposing the Georgi-Jarlskog relations 
\cite{gj} at $\Lambda_{GUT}$, $m^0_s \simeq m^0_\mu/3,\ m^0_d \simeq 3m^0_e$
with Yukawa coupling unification holding for $\tan \beta \sim 5$.
The matrix element contributions can be neatly understood in terms of 
Froggatt-Nielsen diagrams \cite{fn}.  In particular, ``$1's$'' are obtained 
from the ${\bf 16_3 \cdot 16_3 \cdot 10_H}$ vertices; the ``$\epsilon$'' 
terms are obtained from diagrams exhibiting the $\langle {\bf 
45_H}\rangle_{B-L}$ suppression;
while the ``$\sigma$'' terms arise from the ${\bf 16_2 \cdot
16_H \cdot 16'_H \cdot 16_3}$ effective operator which contributes
only to $D$ and $L$ in the lop-sided fashion indicated.  The other entries
arise from more complex diagrams \cite{ab}.

All nine quark and charged lepton masses plus the three CKM angles and CP 
phase are well-fitted with the eight input parameters.  The vertex of the 
CKM unitary triangle occurs at the center of the presently allowed region
with $\sin 2\beta \simeq 0.65$.  The Hermitian matrices $U^\dagger U,\ 
D^\dagger D$, and $N^\dagger N$ are diagonalized with small left-handed 
rotations, while $L^\dagger L$ is diagonalized by a large left-handed rotation.
This neatly accounts for the fact that $V_{cb} = (U^\dagger_U U_D)_{cb}$ is 
small, while $U_{\mu 3} = (U^\dagger_L U_\nu)_{\mu 3}$ is large for any 
reasonable $M_R$.

Since the solar and atmospheric mixings are essentially decoupled in the 
model, the structure of the right-handed Majorana mass matrix determines the 
type of $\nu_e \leftrightarrow \nu_\mu,\ \nu_\tau$ solar neutrino mixing. 
\begin{itemize}
        \item   The {\bf SMA Solar Neutrino Solution} can be obtained with 
        \begin{equation}
	  M_R = \left(\matrix{C & 0 & 0\cr 0 & B\epsilon^2 & 0\cr
                        0 & 0 & 1\cr}\right)\Lambda_R\\
	\end{equation}
                and $B = -1.9,\ C = 5\times 10^{-8},\ \Lambda_R = 1.2 \times 
                        10^{14}$ GeV; for small mixing in the 1-2 sector
		of the matrices arises since $N$ and $L$ are nearly 
		diagonal there.
        \item   The {\bf QVO Solar Neutrino Solution} can be obtained with
        \begin{equation}
	  M_R = \left(\matrix{0 & A\epsilon^3 & 0\cr 
                        A\epsilon^3 & 0 & 0\cr
                        0 & 0 & 1\cr}\right)\Lambda_R\\
	\end{equation}
                and $A = 0.05,\ \Lambda_R = 2.4 \times 10^{14}$ GeV,
                which leads to a pair of pseudo-Dirac neutrinos.
	\item	The {\bf LMA Solar Neutrino Solution}, unlike the others, 
		requires fine-tuning for $M_R$ and a nearly hierarchical 
		texture:
	\begin{equation}
          M_R = \left(\matrix{c^2 \eta^2 & -b\epsilon\eta & a\eta\cr 
                -b\epsilon\eta & \epsilon^2 & -\epsilon\cr
                a\eta & -\epsilon & 1\cr}\right)\Lambda_R\\
	\end{equation}
		in terms of parameters $\epsilon$ and $\eta$ introduced
		in the Dirac sector.  Note that the 2-3 subsector has 
		zero determinant and is closely related to that of $N$,
		as can also be understood in terms of Froggatt-Nielsen
		diagrams.  

	With $a=1,\ b=c=2$ and $\Lambda_R = 2.5 \times 10^{14}$ GeV,
	by the seesaw mechanism the light neutrino mass matrix becomes
        \begin{equation}
	  M_\nu = \left(\matrix{ 0 & -\epsilon & 0\cr 
                        -\epsilon & 0 & 2\epsilon\cr 0 & 2\epsilon & 1\cr}
                        \right)M^2_U/\Lambda_R\\
	\end{equation}
		with three texture zeros, which leads to
\vspace*{-0.1in}
	\begin{equation} 
        \begin{array}{ll}
          M_1 = M_2 = 2.8 \times 10^{8}\ {\rm GeV},\quad & M_3 = 2.5 
                \times 10^{14}\ {\rm GeV},\\[-0.1in]
          \Delta m^2_{32} = 3.2 \times 10^{-3}\ {\rm eV^2},\quad & 
                \sin^2 2\theta_{\rm atm} = 0.994,\\[-0.1in]
          \Delta m^2_{21} = 6.5 \times 10^{-5}\ {\rm eV^2},
		\quad & \sin^2 2\theta_{\rm sol} = 0.88,\\[-0.1in]
          U_{e3} = -0.014,\quad &\sin^2 2\theta_{\rm reac} = 0.0008,\\[-0.1in]
        \end{array}
	\end{equation}
                which compares favorably with the present S-K best-fit point 
		in the LMA region cited in the introduction.  Note also
		that $\sin^2 2\theta_{\rm atm}$ is extremely close to 
		maximal. In fact, the whole presently-allowed LMA region can 
		be covered with $1.0 \aprle a \aprle 2.5,\ 1.8 \aprle b=c 
		\aprle 5.2$ as shown in Fig. 1, where contours of constant
		$\sin^2 2\theta_{12}$ and $\sin^2 2\theta_{13}$ are 
		plotted.  From this Figure, the advantage of a 
		neutrino factory over a superbeam facility is apparent for
		this model.  Other plots of similar interest have been
		obtained by the author in collaboration with S. Geer 
		\cite{ag}.
\end{itemize}
\vspace*{-0.4in}
\begin{figure}[h]
\centerline{
\includegraphics{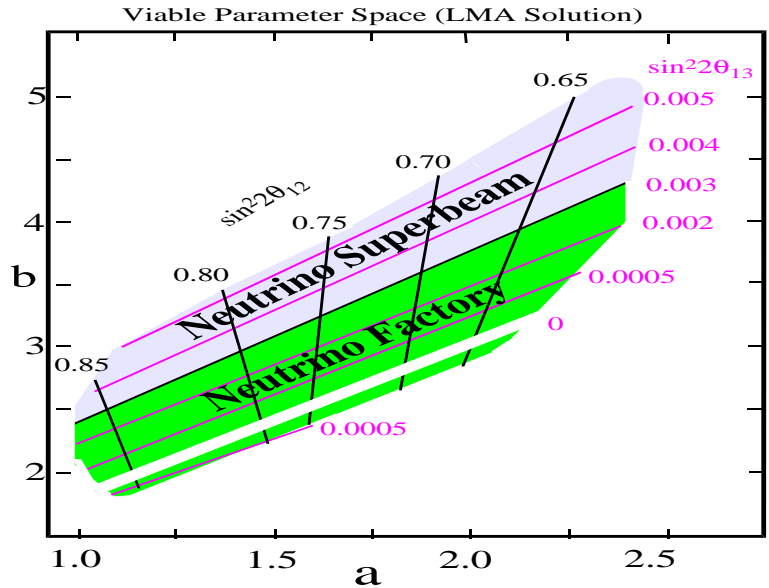}}
\vspace*{4in}
\caption{Input parameter space with $b = c$ vs. $a$ in the right-handed
	Majorana mass matrix for the LMA solution.  Contours of constant
	$\sin^2 2\theta_{12}$ and $\sin^2 2\theta_{13}$ are indicated
	with the reaches of a neutrino superbeam and that for a neutrino
	factory highlighted.  Only the thin sliver around 
	$\sin^2 2\theta_{13} = 0$ corresponding to maximal atmospheric
	neutrino mixing is inaccessible to a neutrino factory.}
\end{figure}
\section{Summary}
\vspace*{-0.2in}
We have seen that $SO(10)$ SUSY GUT models can explain the observed 
atmospheric and solar neutrino oscillation data within the (3 active, 
0 sterile) neutrino framework.  Unfortunately, there is no strong preference 
for any particular solar neutrino solution, though the SMA, QVO and LOW 
solutions are easiest to obtain, with the LMA solution requiring fine-tuning. 
In the model described, that fine-tuning can be understood in terms of 
Froggatt-Nielsen diagrams.  Finally we noted that a neutrino factory is 
essentially required, in order to determine $U_{e3}$ for the present 
fully-allowed LMA region.

\end{document}